\documentclass[aoas,MSNbibl,nameyear,seceqn,dvips]{arximspdf}
\usepackage{dcolumn,url,breakurl}
\usepackage{graphicx}

\doi{10.1214/14-AOAS715} 
\volume{8}
\issue{2}
\pubyear{2014}
\firstpage{974}
\lastpage{998}
\setattribute{copyright}{owner}{In the Public Domain}

\makeatletter
\newcolumntype{d}[1]{D{.}{.}{#1}}
\makeatother

\begin{document}
\begin{frontmatter}

\title{Leveraging local identity-by-descent increases the
power of case/control GWAS with related individuals}
\runtitle{cQLS}

\begin{aug}
\author[a]{\fnms{Joshua N.}~\snm{Sampson}\corref{}\thanksref{nci,t1}\ead[label=e1]{joshua.sampson@nih.gov}},
\author[c]{\fnms{Bill}~\snm{Wheeler}\thanksref{ims}\ead[label=e2]{wheelerb@imsweb.com}},
\author[b]{\fnms{Peng}~\snm{Li}\thanksref{nci}\ead[label=e3]{penglistat@outlook.com}}
\and
\author[a]{\fnms{Jianxin}~\snm{Shi}\thanksref{nci,t1}\ead[label=e4]{jianxin.shi@nih.gov}}
\runauthor{Sampson, Wheeler, Li and Shi}
\thankstext{t1}{Supported by
the intramural program of the National
Institute of Cancer. This study utilized the high-performance
computational capabilities of the Biowulf Linux
cluster at the National Institutes of Health, Bethesda, Md.
(\protect\url{http://biowulf.nih.gov}).}
\affiliation{National Cancer Institute\thanksmark{nci} and Information
Management Services\thanksmark{ims}}
\address[a]{J.~N. Sampson\\
J. Shi\\
Division of Cancer Epidemiology and Genetics\\
National Cancer Institute\\
9609 Medical Center Drive\\
Rockville, Maryland 20850\\
USA\\
\printead{e1}\\
\phantom{E-mail:\ }\printead*{e4}}

\address[b]{P. Li\\
Genetic Epidemiology Group (GEP)\\
International Agency\\
\quad for Research on Cancer (IARC)\\
150 Cours Albert Thomas\\
69008 Lyon\\
France\\
\printead{e3}}

\address[c]{B. Wheeler\\
Information Management Services\\
6110 Executive Blvd.\\
Rockville, Maryland 20852\\
USA\\
\printead{e2}}
\end{aug}

\received{\smonth{8} \syear{2013}}
\revised{\smonth{11} \syear{2013}}

%
\begin{abstract}
Large case/control Genome-Wide Association Studies (GWAS) often include
groups of related individuals with known relationships. When testing
for associations at a given locus, current methods incorporate only the
familial relationships between individuals. Here, we introduce the
chromosome-based Quasi Likelihood Score (cQLS) statistic that
incorporates local Identity-By-Descent (IBD) to increase the power to
detect associations. In studies robust to population stratification,
such as those with case/control sibling pairs, simulations show that
the study power can be increased by over 50\%. In our example, a GWAS
examining late-onset Alzheimer's disease, the $p$-values among the most
strongly associated SNPs in the APOE gene tend to decrease, with the
smallest $p$-value decreasing from $1.23 \times10^{-8}$ to $7.70 \times
10^{-9}$. Furthermore, as a part of our simulations, we reevaluate our
expectations about the use of families in GWAS. We show that, although
adding only half as many unique chromosomes, genotyping affected
siblings is more efficient than genotyping randomly ascertained cases.
We also show that genotyping cases with a family history of disease
will be less beneficial when searching for SNPs with smaller effect sizes.
\end{abstract}

%

\end{frontmatter}

\section{Introduction}
\label{sec1}

Genome-Wide Association Studies (GWAS) of binary traits can include
related individuals from known pedigrees [\citet
{Barrett2008,Willer2008}]. In case--control studies, GWAS generally
increase power by collecting cases with affected relatives, and may
therefore find it cost-effective to genotype these related cases as
well. Furthermore, GWAS may target families because the appropriate
association tests are robust to population stratification [\citet
{Ewens2008}], the families have previously been collected for linkage
analyses, the genetic variants can be called more accurately [\citet
{Wang2007}], or the effects of parental imprinting can be evaluated
[\citet{Wilkinson2007}].

There are two general approaches for evaluating GWAS with related
individuals from known pedigrees [\citet{Ott2011}, Manichai\-kul et~al. (\citeyear{Manichaikul2012})]. The
first approach combines two independent tests, a family-based test
[\citet{Sham2002,Lange2003,L2000}], such as the
transmission-disequilibrium test, and a population-based test of
association. Methods exist for combining these two tests so the
resulting analysis is robust to population stratification [\citet
{Zheng2010,Won2012,Mirea2012,Manichaikul2012}]. The second approach,
which potentially sacrifices robustness for improved power, calculates
a single test statistic of association that accounts for the correlated
genotypes among relatives [\citet
{Bourgain2003,Thornton2007,Slager2001}]. Our focus is on the second
approach and we derive a single test statistic which can use principal
components [\citet{Price2006}] to correct for population stratification
when necessary.

Until now, standard GWAS analyses have focused on the individual. For
each individual, a study records their genotype and disease status. If
all individuals were unrelated, the appropriate test statistic would
simply be the correlation between an individual's genotype and disease
status [\citet{Hirschhorn2005}]. However, when some individuals are
related, a test statistic should correct for the dependence among
genotypes and allow for the minor allele frequency (MAF) at a causal
SNP to be higher in controls with affected relatives, as compared to
randomly ascertained controls [\citet
{Thornton2007,Bourgain2003,Zhu2012}]. The more-powerful
Quasi-Likelihood Score (mQLS) statistic [\citet{Thornton2007}], an
extension of the QLS [\citet{Bourgain2003}], accomplishes both
objectives. However, because these statistics consider only the overall
Identity By Descent (IBD) status between two individuals, they cannot
allow for the MAF at a causal SNP to be higher in controls that share
both alleles (IBD${}={}$2) with an affected sibling, as compared to controls
that share no alleles (IBD${}={}$0) with an affected sibling.

We introduce a GWAS analysis that takes a different perspective and
focuses on the founder chromosomes within each family. For each founder
chromosome, we effectively identify its allele (at a given SNP) and the
proportion of individuals carrying the SNP from that chromosome who are
affected. This step is made possible by recent advances in IBD mapping
and haplotyping [\citet{Browning2010,Peters2012,He2013}]. We suggest a
chromosome-based Quasi Likelihood Score (cQLS) statistic that, at its
simplest, measures the correlation between a binary indicator for the
minor allele and the proportion of individuals affected. Formally, this
test statistic is a partial score statistic from the retrospective
likelihood that includes local IBD status among the observed data. The
cQLS effectively leverages local IBD to improve power in GWAS with a
large number of family-based controls. Furthermore, the derivation of
the cQLS as a score statistic shows how to appropriately handle
families of arbitrary pedigrees, include phenotype data from
ungenotyped relatives, accommodate covariates, and allow for arbitrary
models (e.g., logistic, liability threshold) linking disease risk and
genotype status. In addition to these useful features, the cQLS also
permits permutation-based measures of statistical significance.
Accounting for relatedness in permutation methods of subject-centric
approaches has proven exceptionally difficult [\citet{Wang2011}].

In the next section, we define cQLS, describe the simulated data sets
for testing its performance, and introduce a GWAS of Late-Onset
Alzheimer's Disease (LOAD) conducted by the National Institute of Aging
and the National Cell Repository for Alzheimer's Disease
(NIA-LOAD/NCRAD) [\citet{Lee2008,Wijsman2011}]. In the third section, we
evaluate the performance of the cQLS in simulated data sets and the
NIA-LOAD/NCRAD data set. In addition to showing the potential benefit
of cQLS, simulations evaluate the power gained by genotyping affected
siblings, as compared to randomly ascertained cases, and demonstrate
the diminished benefit of recruiting cases with a family history of
disease when searching for SNPs with small effects [\citet
{Ionita-Laza2011,Hattersley8,Teng1999}]. In the final section, we
conclude with a brief discussion.

\section{Methods}

\subsection{cQLS: Definition}

We consider a case--control study that contains $N_{\mathrm{Fam}}$ families,
labeled $j=1,\ldots,N_{\mathrm{Fam}}$. Within family $j$, all genotyped chromosomes
are assumed to arise from a family-specific set of $n_j$ founder
chromosomes, labeled $k=1,\ldots,n_j$. We denote the total number of
chromosomes by $N_T$:
%
\begin{equation}
N_T = \sum_{j} n_j.
\end{equation}
For a given SNP, we let $Y_{jk}=1$ if the unique founder chromosome $k$
in family $j$ has a minor allele, and $Y_{jk}=0$ otherwise. Note that
the subscripts ``$j_1 k$'' and ``$j_2 k$'' refer to different founder
chromosomes from different families. For our discussion here, we will
assume that $Y_{jk}$ is uniquely identifiable given the observed
genetic data and that we can identify, with certainty, those
individuals in family $j$ who inherited the SNP from founder chromosome
$k$. Both assumptions will be relaxed in the \hyperref[app]{Appendix}. Furthermore, we
will let $T_{jik'k}=1$ if the $k'$th copy ($k' \in\{1,2\}$) of the
specified SNP in individual $i$ is descended from founder chromosome~$k$,
and $T_{jik'k}=0$ otherwise. We let $A_{ji}=1$ if individual $i$
is affected, and $A_{ji}=0$ otherwise.

For each founder SNP, we have $Y_{jk}$, indicating the presence or
absence of a minor allele, and $C_{jk} = \sum_{i,k'}T_{jik'k}A_{ji}$, a
count of the number of affected individuals with that SNP. If we expect
that the minor allele increases disease risk, then the presence of a
minor allele should be associated with a larger number of affected
individuals. Therefore, the squared correlation between $Y_{jk}$ and
$C_{jk}$ should be high. The proposed test statistic is a variation of
this squared correlation.

Instead of using the count, $C_{jk}$, we use a version that normalizes
each individual's affection status to their expected affection status
under the null hypothesis of no association,
%
\begin{equation}
\label{piEq} Z_{jk} = \sum_i \sum
_{k'} T_{jik'k} \biggl(A_{ji}-
\frac{\hat{\pi}_{0}}{1-\hat
{\pi}_{0}}(1-A_{ji})\biggr),
\end{equation}
where $\hat{\pi}_{0}$ is our estimate of the prevalence of the disease
in the population. We presume there are established estimates of $\hat
{\pi}_{0}$ available in the literature. In the \hyperref[app]{Appendix}, we show how we
can modify the expected affection status when other characteristics are
known. We further show how we can make specific individuals, such as
those with a family history of disease, carry more weight in the
analysis by effectively increasing their contribution to $Z_{jk}$. We
denote the average of all observed values of $Z_{jk}$ by
%
\begin{equation}
\bar{Z} \equiv\frac{\sum_{j,k} Z_{jk}}{N_T}.
\end{equation}
Finally, we define a normalized version of the allele for chromosome $k$,
%
\begin{equation}
Y^{\dagger}_{jk}=Y_{jk}-\hat{\phi},
\end{equation}
where
%
\begin{equation}
\hat{\phi} \equiv\frac{\sum_{j,k} Y_{jk}}{N_T}
\end{equation}
and is our estimate of $\phi$, the minor allele frequency in the
population, under the null hypothesis. As promised, our
chromosome-based Quasi-Likelihood Score (cQLS) statistic is then
proportional to the squared correlation between $Y^{\dagger}$ and $Z$,
%
\begin{equation}
\mathrm{cQLS} \equiv\frac{(\sum_{j,k} (Z_{jk}-\bar{Z}) Y^{\dagger}_{jk})^2 } {
\sum_{j,k} (Z_{jk}-\bar{Z})^2 \hat{\phi}(1-\hat{\phi})}. \label{ss}
\end{equation}
Because $\operatorname{var}(Y^{\dagger}_{jk})$ is calculated as $\hat{\phi}(1-\hat{\phi
})$, equation (\ref{ss}) requires the assumption of the Hardy--Weinberg
Equilibrium (HWE). Therefore, the version of cQLS defined by equation
(\ref{ss}) is appropriate in an ideal scenario where the genotypes and
phenotypes of all individuals are known, $Y_{jk}$ can be identified for
all families and chromosomes, and HWE holds. In the \hyperref[app]{Appendix}, we define
a more robust cQLS that allows for violations in all three assumptions.

\subsection{Assigning chromosomes}

To calculate cQLS, we need to determine $T_{jik'k}$, or the identity of
the two chromosomes in individual $i$ for all $i,j,k'$, and $k$ at each
SNP. We perform this calculation in three steps. First, we phase
subjects using BEAGLE [\citet{Browning2011}], a software package for
analysis of large-scale genetic data sets. Second, we detect shared
segments within a family using GERMLINE [\citet{Gusev2009}], a software
package for discovering long shared segments of Identity-By-Descent
(IBD). In the third step, we convert the IBD status to $T_{jik'k}$. The
algorithm for this final step is described in Appendix \ref{algo}.

\subsection{Simulations}

Our three aims are to (1) assess the benefit of chromo\-some-based
association tests (2) assess the value of genotyping an affected
sibling and~(3) assess the benefit of genotyping cases known to have
affected siblings.

For our simulations, we assume that the liability threshold model
accurately describes disease risk. In the liability threshold model, a
complex disease results from an underlying, normally distributed,
phenotype, or liability. When an individual's liability exceeds a
specific threshold, the individual is affected. In our scenario, the
liability, $\mathbf{L}_{j}$, for individuals in nuclear family $j$ can
be described by a linear function of their genotypes $\mathbf{G}_{j}$,
where $G_{ji}$ is the number of minor alleles for individual $i$,
$\mathbf{G}_{j} = [
{ G_{j1}\enskip \cdots \enskip G_{jN_j}}
]^t$, and the superscript $t$ indicates transpose. Here, $\mathbf
{E}_{j}$ accounts for environmental factors, while $\mathbf{F}_{j}$
accounts for background genetic correlation:
%
\begin{equation}
\mathbf{L}_{j}=\beta_G (\mathbf{G}_{j}-2
\bolds{\phi}) + \mathbf{E}_{j} + \beta_F
\mathbf{F}_{j},
\end{equation}
where
%
\begin{equation}
\left[\matrix{ E_{j1} \vspace*{2pt}
\cr
E_{j2}
\vspace*{2pt}
\cr
\vdots\vspace*{2pt}
\cr
E_{jN_j} }\right] %
\sim N
\left( %
\left[\matrix{ 0 \vspace*{2pt}
\cr
0\vspace*{2pt}
\cr
\vdots
\vspace*{2pt}
\cr
0 }\right] %
, %
\left[\matrix{ 1 & 0 & \cdots& 0
\vspace*{2pt}
\cr
0 & 1 & \cdots& 0 \vspace*{2pt}
\cr
\vdots\vspace*{2pt}
\cr
0 & 0 &
\cdots& 1 }\right] %
\right)
\end{equation}
and
%
\begin{equation}
\left[\matrix{ F_{j1} \vspace*{2pt}
\cr
F_{j2}
\vspace*{2pt}
\cr
F_{j3} \vspace*{2pt}
\cr
\vdots\vspace*{2pt}
\cr
F_{jN_j}}\right ] %
\sim N\left( %
\left[\matrix{ 0 \vspace*{2pt}
\cr
0\vspace*{2pt}
\cr
0\vspace*{2pt}
\cr
\vdots\vspace*{2pt}
\cr
0 }\right] %
,
\left[\matrix{ 1 & 0 & 0.5 & \cdots& 0.5 \vspace*{2pt}
\cr
0 & 1 & 0.5 &
\cdots& 0.5 \vspace*{2pt}
\cr
0.5 & 0.5 & 1 & \cdots& 0.5\vspace*{2pt}
\cr
\vdots&
\vdots& \vdots& \ddots& \vdots\vspace*{2pt}
\cr
0.5 & 0.5 & 0.5 & \cdots& 1 }\right]
\right).
\end{equation}
Here, $N_j$ is the number of individuals in family $j$, and individuals
$j1$ and $j2$ are the parents. The liability threshold model will
assign all individuals with $L_{ji}$ exceeding the population's
95th percentile to have the disease. We chose $\beta_G$ so that
the corresponding OR for each additional minor allele was between 1.02
and 1.36. We chose $\beta_F$ so that the sibling recurrence risk ratio,
$\lambda_S$, was 1.5, 2 or 5 when $\beta_G=0$. Note that the magnitude
of the sibling recurrence risk ratio, as defined here, is independent
of the strength of the tested SNP. To calculate $\beta_F$, we
numerically solved the following equation:
%
\begin{equation}
\int_{F_2^{-1}(0.95)}^{\infty} \biggl[\int_{F_2^{-1}(0.95)}^{\infty}
f_1(x|y)\, \partial x\biggr] f_2(y) \,\partial y = 0.05
\lambda_S,
\end{equation}
where $F_2$ and $f_2$ are the cumulative distribution and density of
$N(0,1+\beta^2_F)$, and $f_1(x|y)$ is the density of $N(\beta^2_F
y/(2(1+\beta^2_F)),(1+2\beta^2_F+1.75\beta^4_F)/(1+ \beta^2_F))$. The
MAF in the population was fixed at 0.1, and genotypes were simulated
under HWE.

Our simulated studies, summarized in Table~\ref{simStudy}, collect
families and/or unrelated cases and controls. The first five studies
all start by identifying and genotyping 10,000 unrelated, randomly
ascertained controls (R.A. controls). By randomly ascertained, we mean
that we have no knowledge of their family history of disease. Study 1
further genotypes 5000 unrelated, randomly ascertained cases (R.A.
cases). Study 2 genotypes 5000 unrelated cases (F.H. cases) with a
family history of disease, specifically with an affected sibling. Study
3 genotypes both the 5000 R.A. cases and the 5000 F.H. cases. Study 4
genotypes 10,000 R.A. cases and 5000 F.H. cases. Studies 5 and 6
genotype 1 sibling of each proband. Study 5 genotypes the 5000 R.A
cases, 5000 F.H. cases and their affected siblings. Study 6 includes
10,000 unrelated cases with an unaffected sibling and their unaffected
siblings. Study 7 genotypes multiple relatives of each proband.
Specifically, study 7 genotypes the 10,000 F.H. cases, their affected
siblings and two unaffected siblings. Studies 6 and 7 highlight the
potential power gain for cQLS in scenarios where only family-based
controls are available.

\begin{table}
\caption{Each evaluated study design, labeled 1 through 7, genotypes
different types of cases and controls. The first set of columns lists
the number of randomly ascertained cases (Random), identified cases---or
probands---with a family history (Fam. His.), and affected siblings of
probands (Sib. 1) genotyped for the given study. The second set of
columns lists the number of randomly ascertained controls (Rand.),
unaffected ``older'' siblings (Sib. 1) and unaffected ``younger'' siblings
(Sib. 2) genotyped for the given study. Studies 1 and 2 are used to
assess the benefit of genotyping individuals with a family history of
disease. Studies 3, 4 and 5 are used to assess the benefit of
genotyping siblings of probands. Studies 6 and 7 are used to assess
chromosome-based tests}
\label{simStudy}
\begin{tabular*}{\textwidth}{@{\extracolsep{\fill}}lcccccc@{}}
\hline
 & \multicolumn{3}{c}{\textbf{Cases}} & \multicolumn{3}{c@{}}{\textbf{Controls}}\\[-6pt]
& \multicolumn{3}{c}{\hrulefill} & \multicolumn{3}{c@{}}{\hrulefill} \\
\textbf{Study}& \textbf{Random} & \textbf{Fam. His.} & \textbf{Sib. 1} & \textbf{Random} & \textbf{Sib. 1} & \textbf{Sib. 2} \\
\hline
1&\phantom{0.}5000&\phantom{0000.}0&\phantom{0000.}0&10,000&\phantom{0000.}0&\phantom{0000.}0\\
2&\phantom{0000.}0&\phantom{0.}5000&\phantom{0000.}0&10,000&\phantom{0000.}0&\phantom{0000.}0\\
3&\phantom{0.}5000&\phantom{0.}5000&\phantom{0000.}0&10,000&\phantom{0000.}0&\phantom{0000.}0\\
4&10,000&\phantom{0.}5000&\phantom{0000.}0&10,000&\phantom{0000.}0&\phantom{0000.}0\\[3pt]
\multicolumn{7}{c}{Multiple family members genotyped}\\
5&\phantom{0.}5000&\phantom{0.}5000&\phantom{0.}5000&10,000&\phantom{0000.}0&\phantom{0000.}0\\
6&\phantom{0000.}0&10,000&\phantom{0000.}0&\phantom{0000.}0&10,000&\phantom{0000.}0\\
7&\phantom{0000.}0&10,000&10,000&\phantom{0000.}0&10,000&10,000\\
\hline
\end{tabular*}
\end{table}

For various combinations of $\beta_G$, $\lambda_S$ and study design, we
simulated 10,000 data sets and calculated the FBAT test statistic [\citet
{L2000}], mQLS [\citet{Bourgain2003,Thornton2007}] and cQLS for each data
set. The power was defined as the proportion of $p$-values that were
below the $10^{-7}$ threshold, a common threshold for GWAS. By using
10,000 data sets, the standard errors for our estimates of power are
bounded by 0.01. A short description of FBAT and mQLS are provided in
Appendix \ref{mQLSFBAT}. Previous comparisons [\citet{Manichaikul2012}]
of the available test statistics have demonstrated that, among
currently available options, mQLS is consistently a top performer and
the appropriate reference for comparison. For all simulations, the cQLS
statistic was calculated using the true IBD of all chromosomes within
each family. Moreover, the tested SNP was assumed to be the causal SNP.
We also evaluated the effect of mistakes in IBD assignment on the power
of the cQLS statistic. For a proportion of sibling sets, we randomly
assigned the IBD status for each individual in that set, ensuring only
that the IBD status was consistent with the observed genotypes. We then
recalculated the power for Studies 6 and 7 with $\lambda_S=1.5$. We
examined the scenarios where 2\%, 5\% and 8\% of sibling sets were
allowed to have errors.

To test the accuracy of the $p$-value, we simulated $10^7$ data sets
under the null distribution for study designs 5, 6 and 7, where the
mQLS and cQLS statistics differ, assuming $\lambda_S=5$. For
computational efficiency, we included only 1000 subjects and only
examined larger thresholds of $10^{-3}$ and $10^{-4}$.

\subsection{NIA-LOAD}

The details of the NIA-LOAD/NCRAD GWAS of Late Onset Alzheimer's
Disease (LOAD) have been described elsewhere [\citet{Wijsman2011}].
Briefly, the study recruited families with multiple affected
individuals. Specifically, probands were required to have a diagnosis
of Alzheimer's Disease after the age of 60, have a sibling with a
similar diagnosis and an additional biologically-related family member
available for genotyping. In these families, additional relatives over
the age of 50 were recruited regardless of cognitive status. Study
participants were then genotyped using Illumina's Human610Quadv1B
BeadChips (Illumina, San Diego, CA, USA). We augmented the genotypes by
imputing SNPs in the APOE region that have been previously associated
with LOAD [\citet{Bertram2007}] using IMPUTE2 software version 2.2.2
[\citet{Howie2009}], with prephasing by SHAPEIT software version~1 [\citet
{Delaneau2013}] and version 3 of the 1000 Genomes Project data as the
reference set. At each imputed SNP, we assigned the most likely
genotypes that were consistent with IBD status. In order to create a
study guaranteed to be robust to population stratification, we focused
on 115 sets of siblings and pruned each set so that it had an equal
number of cases and controls.

\section{Results}

\subsection{cQLS vs mQLS}

\begin{figure}[t!]

\includegraphics{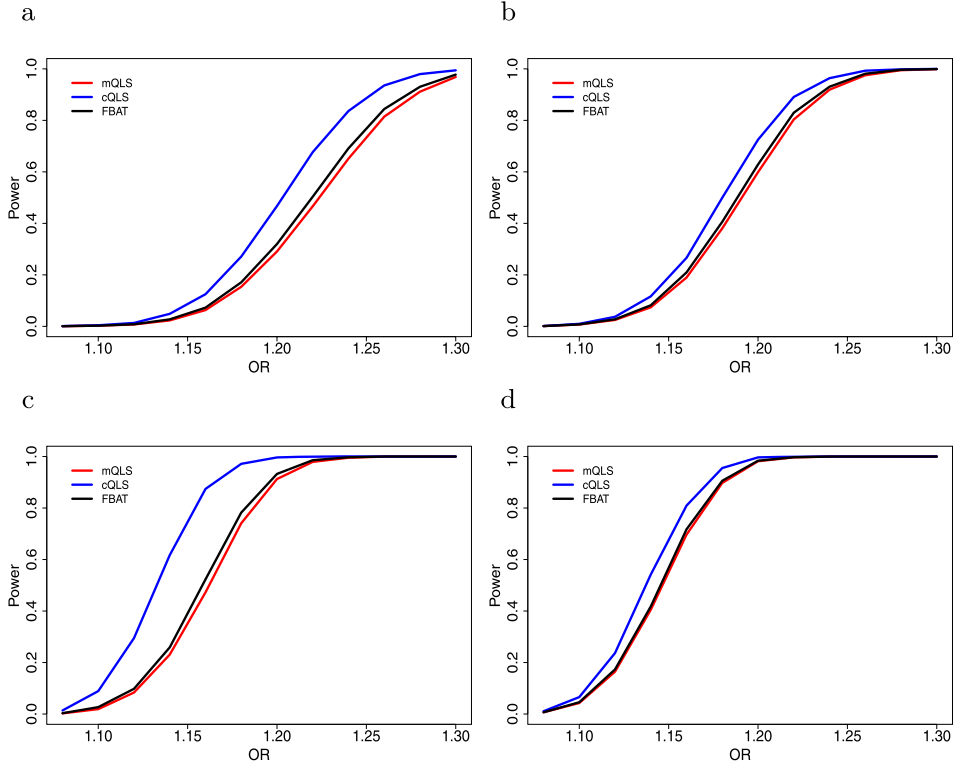}

\caption{The power for cQLS (blue), mQLS (red) and FBAT (black) to
detect the association between a SNP and the disease as a function of
the OR when \textup{(a)} $\lambda_S = 1.5$ in study 6, \textup{(b)}
$\lambda_S = 5$ in
study 6, \textup{(c)} $\lambda_S = 1.5$ in study 7 and \textup{(d)} $\lambda_S = 5$ in
study 7.
Studies 6 and 7
correspond to genotyping pairs of siblings and groups of four siblings,
respectively.
$\lambda_S$ is the sibling relative risk of the disease.}
\label{fig1}
\end{figure}

Chromosome-based test statistics had higher power, as compared to mQLS
and FBAT, to detect associations when the study included cases and
controls from the same family (Figure~\ref{fig1}). We first consider
simulations where the disease had a relatively low sibling relative
risk of $\lambda_S=1.5$. If the GWAS included pairs of siblings (study
6), one affected and one unaffected, then a SNP that would be detected
by cQLS in 75\% of such studies would only be detected by QLS in 51\%
[Figure~\ref{fig1}(a)]. In studies that included sets of four siblings
(study 7), with each set including two affected and two unaffected
individuals, when cQLS provided a power of 0.75, QLS provided a power
of 0.36 [Figure~\ref{fig1}(c)].
When we simulated a disease with a high sibling relative risk ($\lambda
_S=5$), the power gained from using cQLS decreased. For study designs 6
and 7, mQLS achieved a power of 0.59 and 0.63 for SNPs where cQLS
achieved a power of 0.75 [Figure~\ref{fig1}(b), (d)].

Errors in IBD assignment decreased the power for association tests
using cQLS. With an error rate of 2\%, tests based on cQLS still had
higher power for studies 6 and~7. However, with an error rate of 5\%,
cQLS performed no better than the other two test statistics in study
6 ($\lambda_S=1.5$). Specifically, when cQLS had a power of 0.75 (SE${} = {}$0.01),
QLS provided a similar power of 0.72 (SE${} = {}$0.01). When the error
rate reached 8\%, the three test statistics performed similarly in
study 7 ($\lambda_S=1.5$), with QLS achieving a power of 0.74 (0.01)
when cQLS had a power of 0.75 (0.01).

Simulations suggest that the type-I error for the cQLS statistic
matched the chosen $\alpha$ threshold when simulating data from the
null distribution (Table~\ref{genMod}).

%
\begin{table}[t!]
\tabcolsep=0pt
\caption{The proportions of $10^7$ null simulations where the cQLS,
mQLS and FBAT $p$-values are below the specified $\alpha$ threshold. With
$10^7$ simulations, the standard errors for our empirical $\alpha
$-levels of $10^{-3}$ and $10^{-4}$ are approximately
$10^{-5} = \sqrt {10^{-3}/10^{7}}$ and $3 \times10^{-6} = \sqrt{10^{-4}/10^{7}}$}
\label{genMod}
\begin{tabular*}{\textwidth}{@{\extracolsep{\fill}}lcccccc@{}}
\hline
& \multicolumn{2}{c}{\textbf{cQLS}} & \multicolumn{2}{c}{\textbf{mQLS}} & \multicolumn
{2}{c@{}}{\textbf{FBAT}} \\[-6pt]
& \multicolumn{2}{c}{\hrulefill} & \multicolumn{2}{c}{\hrulefill} & \multicolumn
{2}{c@{}}{\hrulefill} \\
\textbf{Study} & $\bolds{\alpha=10^{-3}}$ & $\bolds{\alpha=10^{-4}}$
& $\bolds{\alpha=10^{-3}}$ & $\bolds{\alpha=10^{-4}}$
& $\bolds{\alpha=10^{-3}}$ & $\bolds{\alpha=10^{-4}}$ \\
\hline
5&$9.81 \times10^{-4}$&$9.65 \times10^{-5}$&$9.79 \times
10^{-4}$&$9.03 \times10^{-5}$&$9.44 \times10^{-4}$&$8.38 \times
10^{-5}$\\
6&$9.80 \times10^{-4}$&$9.25 \times10^{-5}$&$1.00 \times
10^{-3}$&$1.08 \times10^{-4}$&$9.56 \times10^{-4}$&$1.04 \times
10^{-4}$\\
7&$9.70 \times10^{-4}$&$1.00 \times10^{-4}$&$9.78 \times
10^{-4}$&$9.68 \times10^{-5}$&$9.62 \times10^{-4}$&$9.28 \times
10^{-5}$\\
\hline
\end{tabular*}   \vspace*{3pt}
\end{table}

\begin{figure}

\includegraphics{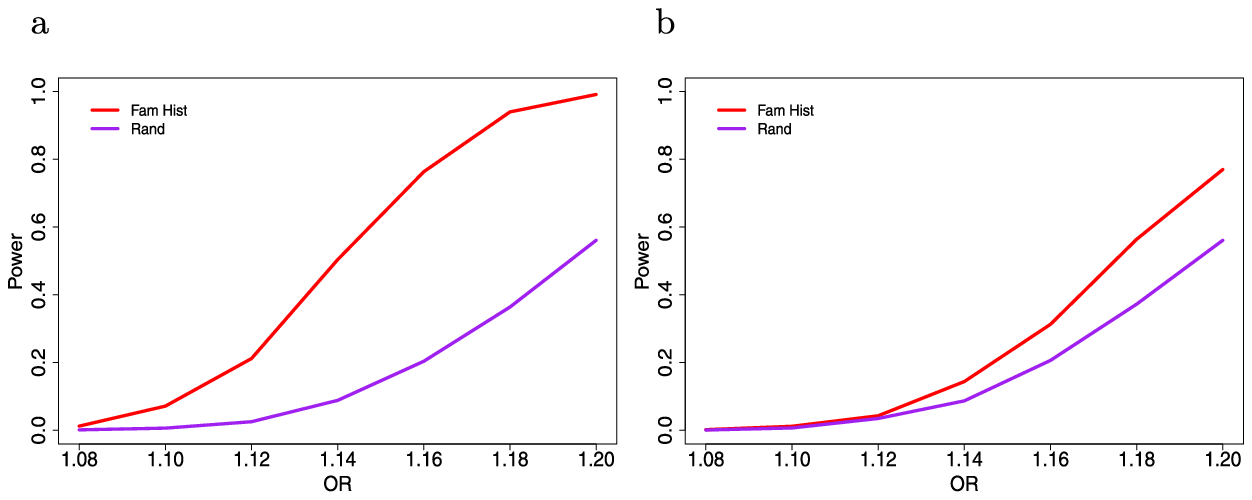}

\caption{\textup{(a)} The power to detect the association between a SNP and the
disease, as a function of OR, when the study compares 10,000 randomly
ascertained controls with either 5000 R.A. cases (red, study 1) or 5000
F.H. cases (blue, study 2) when the sibling relative risk, $\lambda_S$,
is 1.5. \textup{(b)} Same, but with $\lambda_S=5$. For tests of unrelated
individuals, all QLS statistics give identical results.}
\label{fig2}
\end{figure}

\subsection{Genotyping F.H. cases vs R.A. cases}

Genotyping cases with a family history of disease provides a study with
significantly higher power than genotyping randomly ascertained cases.
For a disease with a low sibling relative risk ($\lambda_S=1.5$), when
the effect size for a SNP was large enough so that the study with all
F.H. cases (study 2) had a power of 0.75, a study with all R.A. cases
(study 1) had a power of only 0.20 [Figure~\ref{fig2}(a)]. However, as we
increase the total heritability, shrinking the proportion of
heritability attributable to the tested SNP, the power gained from
using F.H. cases is decreased. When $\lambda_S=5$, a SNP with a power
of 0.75 in study 2 would have had a power of 0.55 [Figure~\ref{fig2}(b)]
in study 1. For interpretation, recall that the sibling relative risk
($\lambda_S$) reflects the heritability from genetic variants other
than the tested SNP. Therefore, as Figure~\ref{fig2}(a) and \ref{fig2}(b)
show, the power of study 1, which collects R.A. cases, does not depend
on $\lambda_S$, but only on the relative risk of the tested SNP.

\begin{figure}

\includegraphics{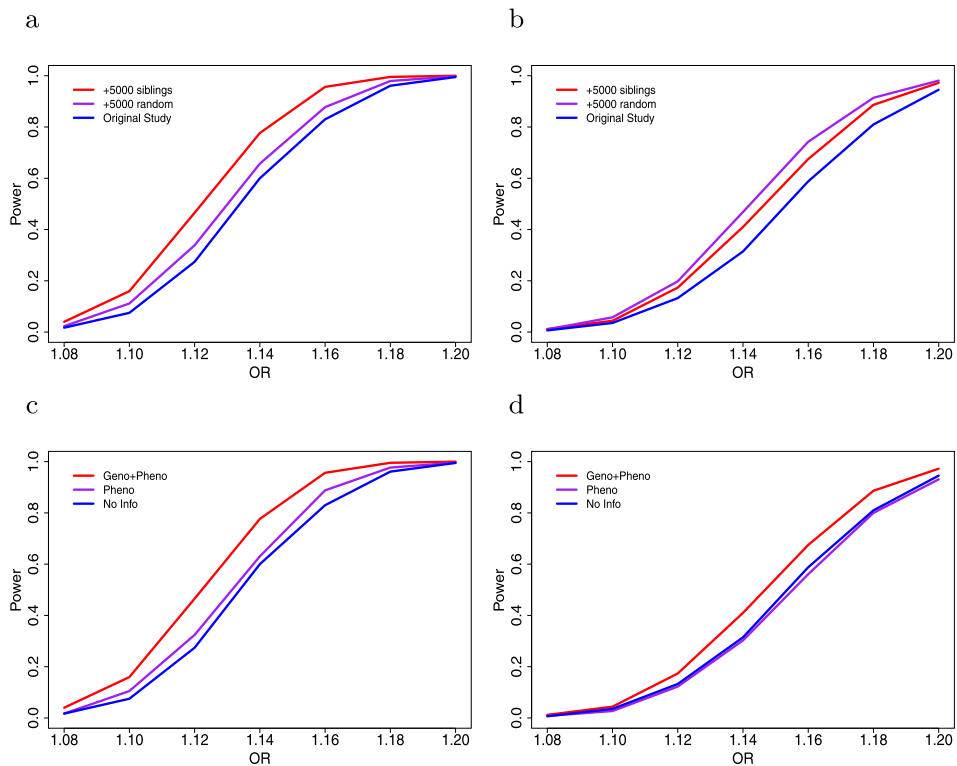}

\caption{\textup{(a)} The power to detect the association between a SNP and the
disease, as a function of OR, when the study compares 10,000 randomly
ascertained controls with either 5000 R.A. cases${} + {}$5000 F.H. cases
(blue, study 3), 10,000 R.A. cases${} + {}$5000 F.H. cases (purple, study 4)
or 5000 R.A. cases${} + {}$5000 pairs of affected siblings (red, study 5)
given a sibling relative risk, $\lambda_S$, of 1.5 and all tests based
on the score statistic. \textup{(b)} Same, but with $\lambda_S=5$.
\textup{(c)} The power
when the study compares 10,000 R.A. controls and either 5000 R.A.
cases${} + {}$5000 F.H. cases and the score statistic (blue), 5000 R.A.
cases${} + {}$5000 F.H.
cases and cQLS (purple), or
5000 R.A. cases${} + {}$5000 pairs of
affected siblings and the cQLS (red) given a sibling relative risk,
$\lambda_S$, of 1.5. \textup{(d)} Same, but with $\lambda_S=5$.}
\label{fig3}
\end{figure}

\subsection{Genotyping affected sibling}

Genotyping the affected siblings of F.H. cases increases the power. The
additional genotyping of 5000 siblings, or moving from study 3 and a
simple score test to study 5 and cQLS, will increase the power to
detect a SNP with OR${}={}$1.15 from 0.57 to 0.75 [Figure~\ref{fig3}(a)]. Each
sibling, on average, only offers one new chromosome, but siblings are
also F.H. cases and, as such, are enriched for the causal allele.
Therefore, genotyping the 5000 siblings promises higher power than
genotyping an additional 5000, unrelated, but randomly ascertained
cases. Moving from study 3 to study 4, which includes 10,000 R.A. cases
and 5000 F.H. cases, only increases the power from 0.57 to 0.63
[Figure~\ref{fig3}(a)]. In studies with a mixture of F.H. cases and R.A. cases,
such as study~3, the power of a standard association test can be
improved by appropriately upweighting F.H. cases through use of either
mQLS or cQLS. When moving from study 3 and cQLS to study 5 and cQLS,
the power gain is less impressive, increasing from 0.60 to 0.75
[Figure~\ref{fig3}(c)]. When the total heritability is high and
$\lambda_S=5$,\vadjust{\goodbreak}
mQLS and cQLS overweight the F.H. cases, and including information
about ungenotyped individuals can potentially lower power [Figure~\ref{fig3}(d)].

\subsection{LOAD}

In addition to the simulations, we tested 11 SNPs in the APOE region of
chromosome 19 for an association with LOAD in participants of the
NIA-LOAD/NCRAD GWAS. Table~\ref{MESA} shows that eight of these 11 SNPs
were associated with LOAD at a $p$-value below 0.05. In 7 of these 8
SNPs, the test based on the cQLS statistic resulted in a lower $p$-value,
as compared to using the mQLS or FBAT statistic. However, all three
methods provide similar evidence that these SNPs are associated with
Alzheimer's disease. For\vadjust{\goodbreak} the two most strongly associated SNPs,
rs429358 and rs4420638, the $p$-values were reduced from $1.23^{-8}$ and
$5.23^{-7}$ based on the mQLS statistic, or $5.9^{-8}$ and $1.2^{-6}$
based on the FBAT statistic, to $7.70^{-9}$ and $1.79^{-7}$ based on
the cQLS statistic.

\begin{table}
\caption{The $p$-values test from association tests of Late-Onset
Alzheimer's Disease and SNPs in the APOE region among the
NIA-LOAD/NCRAD population. The first three columns indicate SNP ID,
chromosome and position. The last three columns indicate the $p$-values
calculated from the mQLS, FBAT and cQLS statistics}
\label{MESA}
\begin{tabular*}{\textwidth}{@{\extracolsep{\fill}}lccccc@{}}
\hline
\textbf{SNP} & \textbf{Chr.} & \textbf{Position} & \textbf{mQLS}
$\bolds{p}$\textbf{-value} & \textbf{FBAT} $\bolds{p}$\textbf{-value} &
\multicolumn{1}{c@{}}{\textbf{cQLS} $\bolds{p}$\textbf{-value}} \\
\hline
rs4806173 & 19 &36,024,925&$8.5\times10^{-1}$&$7.7\times
10^{-1}$&$9.1\times10^{-1}$\\
rs12984928& 19 &36,029,852&$8.5\times10^{-1}$&$7.7\times
10^{-1}$&$9.1\times10^{-1}$\\
rs6857 &19 &45,392,254&$6.0\times10^{-5}$&$5.8\times10^{-6}$&$2.0\times
10^{-5}$\\
rs157582 &19 &45,396,219&$4.7\times10^{-5}$&$4.8\times
10^{-6}$&$1.4\times10^{-5}$\\
rs449647& 19 &45,408,564&$1.9\times10^{-1}$&$2.1\times
10^{-1}$&$7.5\times10^{-2}$\\
rs440446& 19 &45,409,167&$1.8\times10^{-2}$&$2.4\times
10^{-2}$&$2.1\times10^{-2}$\\
rs429358& 19 &45,411,941&$1.2\times10^{-8}$&$5.9\times
10^{-8}$&$7.7\times10^{-9}$\\
rs4420638 &19&45,422,946&$5.2\times10^{-7}$&$1.1\times
10^{-6}$&$1.8\times10^{-7}$\\
rs157580 &19&50,087,106 &$3.9\times10^{-2}$&$3.2\times
10^{-2}$&$1.8\times10^{-2}$\\
rs2075650 &19&50,087,459&$3.0\times10^{-4}$&$7.1\times
10^{-5}$&$1.7\times10^{-4}$\\
rs405509 &19&50,100,676 &$4.5\times10^{-2}$&$2.3\times
10^{-2}$&$4.2\times10^{-2}$\\
\hline
\end{tabular*}
\end{table}

\section{Discussion}

Our primary objective was to introduce the chromosome-based
Quasi-likelihood Score (cQLS) statistic and demonstrate that it can
offer increased power to detect associations in GWAS with related
individuals. Specifically, in studies designed to be robust to
population stratification, such as those including sibling sets equally
divided between cases and controls, statistical power can be increased
by over 50\%. The new statistic can also be applied to less robust
study designs, but, like GWAS with unrelated individuals, would then
require adjusting for population-eigenvectors. The derivation of cQLS
as a partial likelihood shows how to easily adjust for covariates in
both logistic and liability-threshold models.

Although our evaluation has focused on single SNP tests with fixed
thresholds for statistical significance (e.g., $10^{-7}$), cQLS can
offer a key, additional advantage when testing groups of SNPs in
linkage disequilibrium. In GWAS with unrelated individuals, genotypes
can be permuted among individuals to obtain permutation-based measures
of significance. In GWAS with related individuals, standard methods are
not appropriate, as individuals are not independent [\citet{Wang2011}].
However, the founder chromosomes are independent and, therefore, it is
straightforward to apply permutation methods with cQLS in GWAS of
related individuals.

In addition to illustrating the improved power, we designed our
simulations to reevaluate our expectations about the use of families in
GWAS. First, affected siblings are often not genotyped because adding,
on average, only one unique chromosome to the study is thought not to
be worth the cost. However, we show that the power gained from
genotyping an affected sibling can actually exceed the power from
genotyping a randomly ascertained case. Second, it is well known that
studies will have higher power when including cases (F.H. cases) with a
family history of disease. F.H. cases should be enriched with
disease-causing variants. However, we show that when a disease is
highly heritable, the enrichment for any specific disease-causing
variant is weaker. Thus, the benefit from genotyping cases with a
family history of disease is lower, as demonstrated by our comparison
of diseases with $\lambda_S=1.5$ and $\lambda_S=5$.

Like mQLS, cQLS offers the ability to use phenotyped, but not genotyped
family members. Such an advantage could also be gained by imputing the
genotype of such individuals and then performing the GWAS using the
entire population, with appropriate adjustment for the uncertainty
introduced by imputation. However such methods are not easily available
and would still not offer the other benefit of cQLS, identifying local IBD.

Although cQLS requires calculating haplotypes for determining IBD, the
test statistic still focuses on finding associations with single SNPs
as opposed to haplotypes. A haplotype analysis [\citet{akey2001}], which
looks for associations with a specific haplotype, will decrease power
when the causal SNP is directly genotyped, as is likely to be the case
when using dense arrays or sequencing. However, having identified the
haplotypes, this information can be used to adjust for local ancestry
instead of using a more global principal components approach [\citet{wang2011a}].

The cQLS has limitations. First, the statistic will lose power in the
presence of IBD error. In our examples, we found an error rate of 5\%
was large enough to offset the benefit of cQLS in the smaller studies.
A second issue is that this new statistic requires a larger
computational investment. Specifically, the first three steps of
phasing, detecting shared segments of IBD and calculating $T_{jik'k}$
required a total of 8.8 hours on a single 2.8 GHz Intel X5660 processor
for the NIA-LOAD/NCRAD GWAS containing 575003 SNPs. After these initial
steps, calculating the cQLS statistic for the NIA-LOAD/NCRAD GWAS
required only 3.2 minutes. In comparison, mQLS required 70.4 minutes
and FBAT required 11.1 minutes. Both mQLS and FBAT required less time
by dividing the genome into 22 regions. For mQLS, we divided the genome
into 22 intervals with an equal number of SNPs and for FBAT, we divided
the genome into chromosomes. The mQLS program was not designed to
handle GWAS, and we would expect that if optimized, performing an mQLS
analysis would require less computational time than a cQLS analysis.

The benefit of cQLS depends on study design. For those designs that
include mixtures of family-based and randomly ascertained controls, the
increased power offered by cQLS will be lowered. Therefore, the
additional computational cost would offer less value. Second, we have
examined cQLS only in nuclear families and simple three-generation
families (data not shown) where IBD can be reconstructed with high
accuracy. We need further testing to assess the quality of IBD
estimates from more distant relationships. Third, cQLS tests for
association and, unlike FBAT, will have no power when there is linkage
but no association. Although sequencing has removed our reliance on tag
SNPs, a linkage analysis may still offer advantages in the presence of
epistasis.

Finally, we remark that there is no consensus on how to best combine
within-family and between-family information in GWAS with related
individuals. However, a technical way to address this question would be
to examine the full-data likelihood. The derivation of the cQLS
(Appendix \ref{deriv}) starts by defining this likelihood. The form of
the final test statistic is the likelihood ratio test statistic based
on the key partial likelihood, suggesting that the cQLS offers a near
optimal combination of the two types of information.



%


%
%


\begin{appendix}\label{app}
\section{mQLS and FBAT}
\label{mQLSFBAT}
We compared the performance of a cQLS test to two standard tests: mQLS
[\citet{Thornton2007}] and FBAT [\citet{L2000}]. mQLS is a
quasi-likelihood score statistic that presumes an individual's expected
genotype increases with the sum, over all affected family members, of
their kinship coefficients with that individual. A complete, but terse,
definition follows.

For purposes of defining mQLS, we assume there are $n=N_g+M$ total
subjects, of which $N_g$ have been genotyped. We let $\Phi$ be the
$(N_g+M) \times(N_g+M)$ matrix of kinship coefficients and $\Phi
_{N,M}$ be the submatrix containing the last $M$ columns of the first
$N_g$ rows. The phenotype data are coded as $A^{\ddag}$, a column
vector of length $N_g+M$ having $i$th entry 1 if individual $i$ is
affected, 0 if unknown and $\hat{\pi}_0/(1-\hat{\pi}_0)$ otherwise.
$A^{\ddag}_N$ is the vector containing the first $N_g$ elements (e.g.,
genotyped individuals) and $A^{\ddag}_M$ contains the last $M$
elements. Let $G$ be the vector of observed genotypes\vspace*{1pt} ($\frac{1}{2}G_i
\in\{0,0.5,1\}$) for the first $N_g$ individuals. Finally, let~$\mathbf
{1}$ be a column vector of 1's. Then mQLS is defined as
%
\begin{equation}
\mathrm{mQLS} = \hat{\sigma}^{-2} \bigl(\tfrac{1}{2} G - \hat{
\mu}_0\bigr)^T \alpha\Gamma ^{-1}
\alpha^T \bigl(\tfrac{1}{2} G - \hat{\mu}_0\bigr),
\end{equation}
where
%
\begin{eqnarray}
\alpha& = & A^{\ddag}_N + \Phi^{-1}
\Phi_{N,M} A^{\ddag}_M,
\\
\Gamma& = & \alpha^T \bigl(\Phi A^{\ddag}_n +
\Phi_{N,M} A^{\ddag}_M\bigr) - \bigl(
\mathbf{1}^T \alpha\bigr)^2\bigl(\mathbf{1}^T
\Phi^{-1} \mathbf{1}^T\bigr)^{-1},
\\
\hat{\mu}_0 & = & \hat{p}_{\mathrm{null}} \mathbf{1},
\\
\hat{\sigma}^{-2}_0 & = & \bigl[\tfrac{1}{2}
\hat{p}_{\mathrm{null}} (1 - \hat {p}_{\mathrm{null}})\bigr]^{-1},
\\
\hat{p}_{\mathrm{null}} & = & \tfrac{1}{2} \bigl(\mathbf{1}^T
\Phi^{-1} \mathbf {1}\bigr)^{-1} \mathbf{1}^T
\Phi^{-1} G.
\end{eqnarray}
For implementation, we downloaded the software from
\url{http://galton.uchicago.edu/\textasciitilde mcpeek/software/MQLS/index.html}
and, in simulations, set the prevalence
of the disease to the ``true'' value.

FBAT compares the genotypes observed in the cases to their expected
value under the null hypothesis of ``no linkage and no association'' or
``no association, in the presence of linkage,'' conditioned on the
parent's genotypes (or the appropriate sufficient statistic if parental
genotypes are unknown). For details, we suggest the user's manual for
the software downloadable from
\url{http://www.biostat.harvard.edu/\textasciitilde fbat/default.html}.

\section{cQLS: Derivation}
\label{deriv}

\subsection{Model assumptions}
\label{A1}

Without loss of generality, we can assume that all $N_{\mathrm{ind}}$
individuals come from the same family, and therefore drop the subscript
$j$ from notation.
We further define $\Theta$ to be the set of parameters in the model, including those defining the SNP's effect on the disease.
All notation and discussion assume a single SNP
under study.

We will assume random mating.

We will assume
\[
P[A_i=1|G_i=2,\mathbf{X}_{i},\Theta] =
H_{i}^2 P[A_i=1|G_i=0,\mathbf
{X}_i \Theta],
\]
where $G_i$ is defined as the genotype, or the number of minor alleles,
for individual~$i$, $\mathbf{X}_i$ is a vector of covariates, $\Theta$
is a set of parameters, and
%
\begin{equation}
H_i = \frac{P[A_i=1|G_i=1,\mathbf{X}_i]}{P[A_i=1|G_i=0,\mathbf{X}_i]}.
\end{equation}

The immediate consequence is that
\[
P[A_i=0|G_i=2,\mathbf{X}_{i},\Theta]=(1-
\pi_i)\biggl[\frac{1-H_i\pi_i}{1-\pi
_i}\biggr]^2+
\frac{\pi_i}{1-\pi_i}(H_i-1)^2,
\]
where $\pi_i$ is the probability individual $i$ is affected.

For purposes of \textit{deriving} our test statistic, we further assume
that $(\pi_i/(1-\pi_i))(H_i-1)^2$ is small or that we can treat the
following approximation as an equality without issue:
\[
P[A_i=0|G_i=2,\mathbf{X}_{i},\Theta]
\approx\biggl(\frac{1-H_i\pi_i}{1-\pi
_i}\biggr)^2 P[A_i=0|G_i=0,
\mathbf{X}_{i}].
\]

The properties of the test statistic, such as being distributed as a
$\chi^2_1$ variable under the null, will not depend on this
approximation holding.

These approximations give us two simplifying results:
%
\begin{eqnarray}
 P\bigl[Y_1,\ldots,Y_{n}|\mathbf{A},\mathbf{X},M^*,
\Theta\bigr]&=&\prod_{k} P\bigl[Y_k|
\mathbf{A},\mathbf{X},M^*,\Theta\bigr] \label{ind}
\\
 P\bigl[Y_k=1|\mathbf{A},\mathbf{X},M^*,\Theta\bigr] &=&
\frac{p \prod_i h_i^{\sum
_{k'}T_{ik'k}}}{1+p \prod_i h_i^{\sum_{k'}T_{ik'k}}}, \label{theprob}
\end{eqnarray}
where $h_i = H_i A_i + (1-H_i \pi_i)(1-\pi_i)^{-1}(1-A_i)$, $p =
P[Y_i=1]/P[Y_i=0] = \phi/(1-\phi)$ and $M^* = \{T_{ik'k}\dvtx i \in
{1,\ldots,N_{\mathrm{ind}}}, k' \in\{1,2\}, k \in\{1,\ldots,n\} \}$ is the IBD
architecture. We will derive equations (\ref{ind}) and (\ref{theprob})
in a later section of the \hyperref[app]{Appendix}.

\subsection{Probability and score}
\label{score}

We are interested in the distribution of $\mathbf{Y}$ and $\mathbf
{M}^{O}$ given $\mathbf{A}$, $\mathbf{X}$, $S$ and $\Theta$. Here,
$S$ is a vector indicating that the individuals were selected for the study and
$\mathbf{M}^O$ are the observed values of $\mathbf{M}^*$. The variables
$T_{ik'k}$ cannot be identified in family members that are phenotyped,
but not genotyped. We have chosen to treat $Y$ as the outcome because
we will not need to know the selection procedure used to choose
families and we do not need to estimate the nuisance parameter that is
the correlation of disease status in the family due to nongenetic
similarities. We do note that a full probability would be $P[\mathbf
{Y}, \mathbf{M}^{O}, \mathbf{A}, \mathbf{X}|S, \Theta]$ and we ignore
$P[ \mathbf{A}, \mathbf{X}|S,\Theta]$ because it carries little
information about $\Theta$:
\[
P\bigl[\mathbf{Y},\mathbf{M}^{O}|\mathbf{A},\mathbf{X},S,\theta\bigr]
= P\bigl[\mathbf {Y}|\mathbf{A},\mathbf{X},\mathbf{M}^{O},S,\theta
\bigr] P\bigl[\mathbf {M}^{O}|\mathbf{A},\mathbf{X},S,\theta\bigr].
\]
However, we consider only the conditional probability, as this half is
far more sensitive to $\Theta$. Also, although not mentioned above, we
will make the additional assumption that, conditional on all other
information, the genotypes and the characteristics used to select the
individuals are independent. Then we know that $S$ drops out of the
desired probability:
\begin{eqnarray}
P\bigl[\mathbf{Y}|\mathbf{A},\mathbf{X},\mathbf{M}^{O},S,\Theta
\bigr] &=& \frac
{P[\mathbf{Y},S|\mathbf{A},\mathbf{X},\mathbf{M}^{O},\Theta
]}{P[S|\mathbf{A},\mathbf{X},\mathbf{M}^{O},\Theta]}
\nonumber
\\[-8pt]
\\[-8pt]
\nonumber
&= &P\bigl[\mathbf {Y}|\mathbf{A},\mathbf{X},
\mathbf{M}^{O},\Theta\bigr].
\end{eqnarray}




In an upcoming section, we will show that the score statistic for
$P[\mathbf{Y}|\mathbf
{A},\mathbf{X},\break  \mathbf{M}^{O},\Theta]$ is defined by equation (\ref
{ss}) when $M^*$ is known.
\subsection{Mathematical detail}

\subsubsection{Detail for Section \texorpdfstring{\protect\ref{A1}}{B.1}}

We first demonstrate equation (\ref{ind}):\vspace*{-1pt}
%
\begin{eqnarray}
&&P\bigl[Y_1,\ldots,Y_{n_1}|\mathbf{A},\mathbf{X},M^*,
\Theta\bigr]
\\[-1pt]
&&\qquad=\frac{P[\mathbf{A}|Y_1,\ldots,Y_{n_1},\mathbf{X},\Theta] \prod_k
P[Y_k]}{P[\mathbf{A}|\mathbf{X},M^*,\Theta]}
\\[-1pt]
&&\qquad=\frac{P[\mathbf{A}|\mathbf{Y}=0,\mathbf{X},\Theta] \prod_k \prod_i
h_i^{\sum_{k'} G_{ik'} T^{M^*}_{ik'k}} P[Y_k]}{P[\mathbf{A}|M,\mathbf
{X},\Theta]}
\\[-1pt]
&&\qquad=\prod_{k} P[Y_k|\mathbf{A},M,
\mathbf{X},\Theta].\vspace*{-1pt}
\end{eqnarray}

We next demonstrate equation (\ref{theprob}), where we let $\mathbf
{Y}_{-k}$ be the vector of all alleles except for $k$:\vspace*{-1pt}
%
\begin{eqnarray}
\quad&& P\bigl[Y_k=1|\mathbf{A},\mathbf{X},M^{*},\Theta\bigr]
\\[-1pt]
&&\qquad =P\bigl[Y_k=1|\mathbf{A},\mathbf{X},M^*,\Theta,
\mathbf{Y}_{-k}\bigr]
\\[-1pt]
&&\qquad=\frac{P[\mathbf{A}|\mathbf{Y}_{-k},Y_k=1,\mathbf{X},M^*,\Theta]P[Y_k=1]} {
P[\mathbf{A}|\mathbf{Y}_{-k},\mathbf{X},M^*,\Theta]}
\\[-1pt]
&&\qquad=\biggl(
\frac{P[\mathbf{A}|\mathbf{Y}_{-k},Y_k=1,\mathbf{X},M^*,\Theta
]}{P[\mathbf{A}|\mathbf{Y}_{-k},Y_k=0,\mathbf{X},M^*,\Theta]}
\frac{P[Y_k=1]}{P[Y_k=0]}\nonumber\\[-1pt]
&&\hspace*{6pt}\qquad\quad{}\times  P\bigl[\mathbf{A}|\mathbf{Y}_{-k},Y_k=0,\mathbf
{X},M^*,\Theta\bigr]P[Y_k=0]\biggr)\\[-1pt]
&&\qquad\quad{}\Big/ {
P\bigl[\mathbf{A}|\mathbf{Y}_{-k},\mathbf{X},M^*,\Theta\bigr]} \nonumber
\\[-1pt]
&&\qquad=\frac{P[\mathbf{A}|\mathbf{Y}_{-k},Y_k=1,\mathbf{X},M^*,\Theta
]}{P[\mathbf{A}|\mathbf{Y}_{-k},Y_k=0,\mathbf{X},M^*,\Theta]} \frac
{P[Y_k=1]}{P[Y_k=0]} P\bigl[Y_k=0|\mathbf{A},
\mathbf{X},M^*,\Theta\bigr]
\\[-1pt]
&&\qquad=\frac{\prod_{i} h_i^{\sum_{k'}T_{ik'k}} P[A_i|\mathbf
{Y}_{-k},Y_k=0,\mathbf{X},M^*,\Theta]}{\prod_{i} P[A_i|\mathbf
{Y}_{-k},Y_k=0,\mathbf{X},M^*,\Theta]} \frac{P[Y_k=1]}{P[Y_k=0]}
\nonumber
\\[-9pt]
\\[-9pt]
\nonumber
&&\qquad\quad{}\times P\bigl[Y_k=0|\mathbf{A},
\mathbf{X},M^*,\Theta\bigr]
\\[-1pt]
&&\qquad=\prod_{i} h_i^{\sum_{k'}T_{ik'k}}
\frac{P[Y_k=1]}{P[Y_k=0]} P\bigl[Y_k=0|\mathbf{A},\mathbf{X},M^*,\Theta
\bigr].\vspace*{-1pt}
\end{eqnarray}

\subsubsection{Score statistic: $Y_{jk}$ and $M^*$ can be uniquely identified}
\label{YM}

The overall probability can be written as the product of the
probabilities for each chromosome with the assumptions in place:
%
\begin{equation}
P[\mathbf{Y}|D,p,\Theta]= \prod_k
P[Y_{k}|D,p,\Theta],\vspace*{-1pt}
\end{equation}
where $D$ abbreviates the collected data, $D=\{\mathbf{A},\mathbf
{X},M^*\}$.\vadjust{\goodbreak}

Because we must account for the nuisance parameter $p$, the score
statistic for $\Theta$ is equation (\ref{bigmat}) evaluated under the
null hypothesis
%
\begin{eqnarray}\label{bigmat}\qquad
&& - \!\left[\matrix{\displaystyle \sum_{k}
\frac{d}{d\Theta} \ell_D(Y_{k}) & \displaystyle\sum
_{k} \frac{d}{dp} \ell_D(Y_{k})
}\right]
\nonumber
\\[-8pt]
\\[-8pt]
\nonumber
&&\qquad{}\times
\left[\matrix{\displaystyle \sum_{k}
\frac{d}{d^2\Theta} \ell_D(Y_{k}) & \displaystyle\sum
_{k} \frac{d}{d\Theta\,
dp} \ell_D(Y_{k})
\vspace*{2pt}
\cr
\displaystyle\sum_{k} \frac{d}{d\Theta\, dp}
\ell_D(Y_{k})& \displaystyle\sum_{k}
\frac{d}{d^2p} \ell_D(Y_{k}) }\right] %
^{-1} %
\left[\matrix{ \displaystyle\sum_{k}
\frac{d}{d\Theta} \ell_D(Y_{k}) \vspace*{2pt}
\cr
\displaystyle\sum
_{k} \frac{d}{dp} \ell_D(Y_{k})
}\right] %
,
\end{eqnarray}
where
%
\begin{eqnarray}
&\ell_D(Y_{k}) = \log\bigl(P\bigl[Y_{k}|
\mathbf{A},\mathbf{X},M^*,p,\Theta\bigr]\bigr).
\end{eqnarray}

It is straightforward to evaluate the needed derivative
%
\begin{eqnarray}
 U &\equiv&\sum_k \frac{d}{d\Theta}\bigg|_{H_0}
\ell_D(Y_k)= \sum_k
Z_k (Y_k-\hat{\phi}), \label{null3}
\\
 \hat{\phi} &=& \frac{\sum_k Y_k}{\sum_k 1},
\end{eqnarray}
where
%
\begin{eqnarray}
 Z_k &\equiv&\sum_i \sum
_{k'} T_{ik'k} \frac{\dot{h}_i}{h_i},
\\
\dot{h}_i &=& \frac{d}{d \Theta} h_i.
\end{eqnarray}

We can rewrite $U$, so that we can calculate its variance, $\sigma
^2_U$, without computing/inverting the matrix in equation (\ref{bigmat}):
%
\begin{eqnarray}
 U &\equiv&\sum_k \frac{d}{d\Theta}\bigg|_{H_0}
\ell_D(Y_k)= \sum_k
(Z_k-\bar {Z}) (Y_k-\hat{\phi}), \label{null5}
\\
 \bar{Z} &=& \frac{\sum_k Z_k}{\sum_k 1},
\\
 \sigma^2_U &=& 
\phi(1-\phi) \sum
_k (Z_k-\bar{Z})^2 \label{HWEvar}.
\end{eqnarray}

\subsubsection{Score statistic: All individuals are not genotyped and
$M^*$ cannot be uniquely identified}
\label{missInfo}

When all family members are not genotyped, the probability must be
averaged over the $B$ possible IBD states
%
\begin{eqnarray}
&&P\bigl[\mathbf{Y}|\mathbf{A},\mathbf{X},\mathbf{M}^{O},p,\Theta
\bigr]
\nonumber
\\[-8pt]
\\[-8pt]
\nonumber
&&\qquad= \sum_{b=1}^B P\bigl[\mathbf{Y}|
\mathbf{A},\mathbf{X},M^*=m_b,p,\Theta\bigr] P\bigl[m_b|M^O
\bigr].
\end{eqnarray}

We can take advantage of the equality
%
\begin{eqnarray} \label{eqU1}
 &&\frac{d}{d \Theta}\bigg|_{H_0} \log\biggl(\sum
_b c_b P\bigl[\mathbf{Y}| D(m_b),
p, \Theta\bigr]\biggr)
\nonumber
\\[-8pt]
\\[-8pt]
\nonumber
&&\qquad = \frac{d}{d \Theta}\bigg|_{H_0} \sum
_b c_b \log\bigl(P\bigl[\mathbf{Y}|
D(m_b), p, \Theta\bigr]\bigr), 
\end{eqnarray}
where we use the abbreviation $c_b=P[m_b|M^O]$ and use $D(m_b)$ because
$M^*$ is no longer known.

Equation (\ref{eqU1}) shows us that the score for $P[\mathbf{Y}|\mathbf
{A},\mathbf{X},\mathbf{M}^{O},p,\Theta]$ is proportional to the score
we would observe had there been $N_M$ families, where all families had
the observed $\mathbf{Y}$ and $c_bN_M$ of those families had IBD
structure $m_b$. Therefore, equation (\ref{null3}) still holds, so long
as we now let
%
\begin{equation}
Z_k \equiv\sum_i \sum
_{k'} P[T_{ik'k}=1] \frac{\dot{h}_i}{h_i}.
\end{equation}

\subsubsection{Score statistic: $Y_{jk}$ cannot be uniquely identified}
\label{UnY}

In some scenarios, $Y_{k}$ cannot be uniquely identified given the
available genetic information. In these scenarios, we must average the
two possibilities to obtain the value of cQLS. As an example, this
situation occurs in our simulated studies of sibling pairs. When two
siblings have IBD${}={}$1 and are each heterozygous, we cannot determine
whether the shared chromosome has the minor or common allele. We focus
on this specific example to explain the needed adjustment.

Let $Q_j=1$ if the siblings have IBD${} = {}$1 and are both heterozygous.
Furthermore, in such a family, let $Y_1$, $Y_2$ and $Y_3$ denote the
alleles on the chromosome uniquely in the first brother, in both
brothers and uniquely in the second brother, respectively. Let $Z_1$,
$Z_2$ and $Z_3$ be the disease variable for each of those chromosomes.
We know
%
\begin{eqnarray}
&&P\bigl(Q=1| A,X,M^*,p,\Theta\bigr)\\
&&\qquad=
 P\bigl(\{Y_1,Y_2,Y_3\} =\{1,0,1
\}|A,X,M^*,p,\Theta\bigr)
\\
&& \qquad\quad{}+ P\bigl(\{Y_1,Y_2,Y_3\}=\{0,1,0
\}|A,X,M^*,p,\Theta\bigr).
\end{eqnarray}
For families with $Q=1$, we must reevaluate their contribution to the
score equations. Under the null hypothesis, we find
%
\begin{eqnarray}
&&\frac{d \log(P(Q=1|A,X,M^*,p,\Theta))} {
d p} \bigg|_{H_0}
\nonumber
\\[-8pt]
\\[-8pt]
\nonumber
&&\qquad= (2- \phi) \frac{-{\phi^2}/{p^2}}{1-\phi} + (1+\phi)
\frac{\phi}{p^2}
\end{eqnarray}
and
%
\begin{equation}
\frac{d \log(P(Q=1|A,X,M^*,p,\Theta))} {
d \Theta}\bigg|_{H_0}= (1-2\phi) Z_2.
\end{equation}

Our new contributions to the score equations lead us to the following
MLE of~$\phi$:
%
\begin{equation}
\hat{\phi}=\frac{N_1+N_3}{N_0+N_1+N_3},
\end{equation}
where
%
\begin{eqnarray}
N_0 &=& \sum_{j,k} 1(Q_j=0)
1(Y_{jk}=0),
\\
N_1 &=& \sum_{j,k} 1(Q_j=0)
1(Y_{jk}=1),
\\
N_3 &=& \sum_{j,k} 1(Q_j=1).
\end{eqnarray}

Furthermore, we can rewrite $U$ as
%
\begin{equation}
U = \sum_{j,k:Q_j=1} (Z_k-\bar{Z})
(Y_k - \hat{\phi}) + \sum_{j:Q_j=0} (1-2
\hat{\phi}) (Z_2-\bar{Z})
\end{equation}
and the score statistic as
%
\begin{equation}
\frac{U^2}{\hat{\sigma}^2_U},
\end{equation}
where we let $p_u$ be the unique probabilities of each of the eight
possible combinations of $\{Y_1,Y_2,Y_3\}$ when IBD${}={}$1, $Z_{j*}$ be the
possible corresponding contributions from family $j$, and $\hat{\sigma
}^2_U$ be the appropriate estimate of the variance under the null:
%
\begin{eqnarray}
p_{u1} &=& \hat{\phi}^3,
\\
p_{u2} &=& \hat{\phi}^2 (1-\hat{\phi}),
\\
p_{u3} &=& \hat{\phi} (1-\hat{\phi})^2,
\\
p_{u4} &=& (1-\hat{\phi})^3,
\end{eqnarray}
\begin{eqnarray*}
Z_{*j1} &=& \Biggl(\sum_{k=1}^3
(Z_k-\bar{Z}) (1- \hat{\phi})\Biggr)^2,
\\
Z_{*j2} &=&\Biggl(\sum_{k=1}^3
(Z_k-\bar{Z}) (1- \hat{\phi}) - Z_3+\bar{Z}
\Biggr)^2+ \bigl((Z_2-\bar{Z}) (1-\phi)
\bigr)^2
\\
&& {}+\Biggl(\sum_{k=1}^3 (Z_k-
\bar{Z}) (1- \hat{\phi}) - Z_1+\bar{Z}\Biggr)^2,
\\
Z_{*j3} &=& \Biggl(\sum_{k=1}^3
(Z_k-\bar{Z}) (1- \hat{\phi}) - Z_2-Z_3+2
\bar{Z}\Biggr)^2+ \bigl((Z_2-\bar{Z}) (1-\phi)
\bigr)^2
\\
&&{}+\Biggl(\sum_{k=1}^3 (Z_k-
\bar{Z}) (1- \hat{\phi}) - Z_1-Z_2+2\bar{Z}
\Biggr)^2,
\\
Z_{*j4} &=& \sum_{k=1}^3
\bigl((Z_k-\bar{Z}) (- \hat{\phi})\bigr)^2
\end{eqnarray*}
and
%
\begin{equation}
\hat{\sigma}^2_U = \hat{\phi} (1-\hat{\phi}) \sum
_{j,k:\mathrm{IBD}(j) \ne1} (Z_k-\bar{Z})^2 + \sum
_{j:\mathrm{IBD}(j) = 1} \sum_{t=1}^4
p_{ut} Z_{*jt}.
\end{equation}

Accurate haplotyping would overcome this difficulty and allow us
to\break
uniquely identify $Y_{jk}$. As we expect haplotyping to become standard
practice in the very near future [\citet{Peters2012}], we expect that
this step will soon be unnecessary.

\subsubsection{Violation of the Hardy--Weinberg Equilibrium}
\label{violHWE}

Our estimate for the variance of $U$ in equation (\ref{HWEvar}) assumes
that the genotypes are in HWE. As an alternative, start by calculating
the 16 possible values (one for each genotype) of $U= \sum_k (Z_k-\bar
{Z}) (Y_k-\hat{\phi})$ for each family $j$. The second step is to
calculate the probability of each of the 16 genotypes. Given these
probabilities and the possible values of $U$, it is straightforward to
calculate the variance of $U$ for any family under the null hypothesis,
conditional on IBD architecture. Currently, this alternative is only
available for families with at most four founding chromosomes (e.g.,
nuclear families) and, therefore, the remaining goal is to estimate $p
= \{p_{200},p_{110},p_{101},p_{020},p_{011},p_{002}\}$, where $p_{xyz}$
is the probability that the founding individuals include $x$, $y$ and
$z$ individuals with genotypes $G_i=0$, $G_i=1$ and $G_i=2$. We
estimate these six probabilities by effectively maximizing $P(Y|M^O,p)$
with the constraints that $p_{xyz} \ge0$ and that $0.25 p_{110} + 0.5
p_{020} + 0.5 p_{101} + 0.75 p_{011} + p_{002} = \hat{\phi}$.
Specifically, we minimize the following function:
%
\begin{eqnarray}
&& 2\biggl(\sum_j 1(n_j = 2) \biggr)
(M_2 V_2 - A_2)^2 +  3\biggl(
\sum_j 1(n_j = 4) \biggr)
(M_3 V_3 - A_3)^2
\nonumber
\\[-8pt]
\\[-8pt]
\nonumber
&&\qquad{}+ 4\biggl(\sum_j 1(n_j = 4) \biggr)
(M_4 V_4 - A_4)^2,
\end{eqnarray}
where $M_4$ is a $6 \times6$ identity matrix,
\begin{eqnarray*}
M_2& =& %
\left[\matrix{ 1 & 0.5 & 0.25 & 0 & 0 & 0 \vspace*{2pt}
\cr
0 & 0.5 & 0.5 & 1 & 0.5 & 0 \vspace*{2pt}
\cr
0 & 0 & 0.25 & 0 & 0.5 & 1}\right]\quad\mbox{and}\\
 M_3 &=& %
\left[\matrix{ 1 & 0.25 & 0 & 0 & 0
& 0 \vspace*{2pt}
\cr
0 & 0.5 & 0.5 & 0.5 & 0 & 0 \vspace*{2pt}
\cr
0 & 0 & 0.5 &
0.5 & 0.5 & 0 \vspace*{2pt}
\cr
0 & 0 & 0 & 0 & 0.25 & 1 \vspace*{2pt}
\cr
0 & 0.25
& 0 & 0 & 0.5 & 0 }\right]
\end{eqnarray*}
and
$V_2$ is the vector estimating $ \{P(\sum_k Y_k = 0 | n_j = 2),P(\sum_k
Y_k = 1 | n_j = 2),P(\sum_k Y_k = 2 | n_j = 2)\}$, $V_3$ estimates $ \{
P(\sum_k Y_k = 0 | n_j = 3),P(\sum_k Y_k = 1, W = 1 | n_j = 3),P(\sum_k
Y_k = 2 , W = 1| n_j = 3),P(\sum_k Y_k = 3 | n_j = 3),\break  P(\sum_k Y_k = 2,
W = 0| n_j = 3)\}$, $V_4$ estimates
$ \{P(\sum_k Y_k = 0 | n_j = 4),\break P(\sum_k Y_k = 1 | n_j = 4), P(\sum_k
Y_k = 1, G_2 = 1| n_j = 4), P(\sum_k Y_k = 1, G_2 \ne1| n_j = 4),
P(\sum_k Y_k = 3 | n_j = 4), P(\sum_k Y_k = 4 | n_j = 4)\}$, $W$ is a binary
variable indicating whether all alleles are identifiable, and $\{
G_1,G_2\}$ are the founder genotypes.

\subsection{Algorithm for assigning $T_{ijk'k}$}
\label{algo}

We start by arbitrarily assigning numbers to the chromosomes of the
founder individuals and trimming the family so that no two individuals
have IBD${} = {}$2. Founder individuals are defined to be the largest group
possible such that all pairs of founder individuals have IBD${} = {}$0. Let
$A$ be initialized as the founder individuals.

Find an individual, $i_3$, in the compliment of $A$, that meets the
first of the following possible criteria and follow the assignment
mechanism. Add individual $i_3$ to $A$ and then repeat.

\begin{enumerate}[(a)]
\item[(a)] $i_3$ has IBD${} = {}$1 with two individuals in $A$, say, $i_1$ and
$i_2$, that are also IBD${} = {}$1 with each other. Count the number of minor
alleles, among individuals $i_1$, $i_2$ and $i_3$, at all loci in the
shared region. Assume chromosomes in individuals $i_1$ and $i_2$ have
been labeled as $\{1,2\}$ and $\{1,4\}$.
\begin{enumerate}[(a)]
\item[Option (a1).] If (nearly) all counts are even, the chromosomes in
individual $i_3$ are assigned as $\{2,4\}$.

\item[Option (a2).] (Nearly). All counts are not even, and individual $i_3$ is
either IBD${} = {}$0 with all other individuals in $A$ or IBD${} = {}$1 only with
individuals who are IBD${}={}$1 with both $i_1$ and $i_2$. Then the
chromosomes in $i_3$ are labeled as $\{1,X\}$, where $X$ is a new
chromosome number.

\item[Option (a3).] (Nearly). All counts are not even, and individual $i_3$ is
IBD${} = {}$1 with at least two more individuals in $A$, say, $i_4$ and
$i_5$, that are IBD${} = {}$1 with each other, but IBD${} = {}$0 with both $i_1$
and $i_2$. Then the chromosomes in $i_3$ are labeled as $\{1,5\}$,
where we assume the chromosomes in $i_4$ and $i_5$ are labeled as $\{
5,6\}$ and $\{5,8\}$.

\item[Option (a4).] (Nearly). All counts are not even, and individual $i_3$ is
IBD${} = {}$1 with exactly one other individual in $A$, say, $i_4$, that is
IBD${} = {}$0 with both $i_1$ and $i_2$. Then, we label $i_3$ as $\{1,5\}$,
where the chromosomes in individuals $i_4$ have been labeled as $\{5,6\}
$.
\end{enumerate}

\item[(b)] $i_3$ has IBD${} = {}$1 with two individuals in $A$, say, $i_2$ and
$i_3$, that share IBD${} = {}$0 with each other.
\begin{enumerate}[(a)]
\item[Option (b1).] Individual $i_1$ (or $i_2$) has IBD${} = {}$1 with another
individual in A, say, $i_4$, in $A$. Then, we label $i_3$ as $\{2,3\}$,
where the chromosomes in individuals $i_1$, $i_2$ and $i_4$ have been
labeled as $\{1,2\}$, $\{3,4\}$ and $\{1,6\}$.

\item[Option (b2).] Individual $i_1$ and $i_2$ have IBD${} = {}$0 with all other
individuals in~A. Then, we label $i_3$ as $\{1,3\}$, where the
chromosomes in individuals $i_1$ and $i_2$ have been labeled as $\{1,2\}
$ and $\{3,4\}$.
\end{enumerate}
\item[(c)] $i_3$ has IBD${} ={} $1 with only one individual in $A$, say, $i_1$.
\begin{enumerate}[(a)]
\item[Option (c1).] Individual $i_1$ has IBD${} = {}$1 with another individual in
$A$, say, $i_2$. Assign the chromosomes in individual $i_3$ as $\{2,X\}
$, where the chromosomes in individuals $i_1$ and $i_2$ have been
labeled as $\{1,2\}$ and $\{1,4\}$ and $X$ is a new chromosome number.

\item[Option (c2).] Individual $i_1$ has IBD${} ={} $0 with all other individuals in
$A$. Assign the chromosomes in individual $i_3$ as $\{2,X\}$, where the
chromosomes in individuals $i_1$ have been labeled as $\{1,2\}$ and X
is a new chromosome number.
\end{enumerate}
\end{enumerate}

\subsection{Limitations}

The standard method for finding an optimal test statistic starts by
defining the parameter of interest and then writing out the likelihood
of the observed data given this, and possibly other, parameters. In the
GWAS discussed here, such a likelihood would necessarily bridge the
within-family and between-family information, and immediately show how
the two pieces of information should be combined. Here, we have defined
this likelihood and shown that the cQLS is derived as the score
statistic to a specific partial likelihood. However, as that likelihood
shows, we ignore information that can be derived from the observed IBD
structure. For example, if all affected siblings are IBD${}={}$2 at a SNP,
that provides some evidence of an association between SNP and disease.
Although that information is minimal, we are currently looking into
methods for capturing and including this independent information as
well. By using only the partial likelihood, the cQLS is not guaranteed
to result in the most powerful test.
\end{appendix}

\section*{Acknowledgments}
The NIA-LOAD and NCRAD data were downloaded
from dbGaP.

%

%



\printaddresses

\end{document}